\documentclass[twocolumn,amssymb,fleqn]{revtex4} 
\usepackage{epsfig,amssymb,amsmath,graphicx,hyperref  }  
\usepackage{color}

\usepackage{graphicx}
\usepackage[caption=false]{subfig}
\usepackage{multirow} 
\usepackage{tabularx}

\begin{document}

\title{Conformal order and Poincar$\rm{\acute{e}}$-Klein mapping underlying
electrostatics-driven inhomogeneity in tethered membranes}
\author{Honghui Sun and Zhenwei Yao}
\email{zyao@sjtu.edu.cn}
\affiliation{School of Physics and Astronomy, and Institute of Natural
Sciences, Shanghai Jiao Tong University, Shanghai 200240, China}
\begin{abstract} 
Understanding the organization of matter under the long-range electrostatic
  force is a fundamental problem in multiple fields. In this work, based on the
  electrically charged tethered membrane model, we reveal regular structures
  underlying the lowest-energy states of inhomogeneously stretched planar
  lattices by a combination of numerical simulation and analytical geometric
  analysis. Specifically, we show the conformal order characterized by the
  preserved bond angle in the lattice deformation, and reveal the
  Poincar$\rm{\acute{e}}$-Klein mapping underlying the electrostatics-driven
  inhomogeneity. The discovery of the Poincar$\rm{\acute{e}}$-Klein mapping,
  which connects the Poincar$\rm{\acute{e}}$ disk and the Klein disk for the
  hyperbolic plane, implies the connection of long-range electrostatic force and
  hyperbolic geometry. We also discuss lattices with patterned charges of
  opposite signs for modulating in-plane inhomogeneity and even creating 3D
  shapes, which may have a connection to metamaterials design. This work suggests
  the geometric analysis as a promising approach for elucidating the
  organization of matter under the long-range force.
\end{abstract}

\maketitle

\section{Introduction}

The electrostatic force represents an important and versatile interaction to mediate
the organization of materials~\cite{levin2005strange,Leunissen2007}, especially in various
self-assembly processes at the
nanoscale~\cite{Walker2011,kostiainen2013electrostatic} and in extensive
electrolyte solutions~\cite{Levin2002,Holm2001}. For example, the electrostatic
interaction has been exploited by Michael Faraday in the mid-1800s to prepare
colloidal suspensions of charged gold particles that remain stable to the
present day~\cite{edwards2007gold}. The symmetric electrostatic
interaction has shown to provide the symmetry-breaking mechanism for the
formation of a variety of material morphologies in multiple fields, ranging
from the electrostatics-driven chirality~\cite{paci2007chiral,Kohlstedt2007},
the shape transitions of
shells~\cite{vernizzi2007faceting,jadhao2014electrostatics} and
ribbons~\cite{yao2016electrostatics,gao2019electrostatic,zaldivar2020twisting}, and
exceedingly rich conformations of highly charged biomolecules like DNA and
proteins~\cite{gelbart2000dna,cherstvy2011electrostatic,savelyev2011dna},
to the assembly of patterned surface coatings
~\cite{lee2001layer,walker2010self,smoukov2007electrostatically}. Through the
Thomson model~\cite{Thomson1904a,Bausch2003e,mehta2016kinetic,yao2019command}, where the
electrically charged particles are confined on the sphere, and various
generalized versions defined on planar
disk~\cite{berezin1985unexpected,Mughal2007,yao2013topological,soni2018emergent}
and curved surfaces~\cite{bowick2009two}, much has been learned about the crucial role
of electrostatics in packing charged particles on confined geometries. While the
confluence of theory and experiment in the past decades shows that in
confined environments the electrostatic force could create unique topological
defect structures and induce non-Euclidean
geometries~\cite{bowick2009two,nelson2002defects},
the question of how the electrostatic force regulates the organization of
matter in free-standing charged condensed matter systems, which represent a host
of entities in materials science and biology, has not yet been fully
understood.

The goal of this work is to explore the electrostatic phenomenon in the
free-standing, electrically charged tethered membrane from the perspective of
geometry, focusing on the regular
structure underlying the lowest-energy particle configuration as a vestige of
the convoluted relaxation process. The membrane model consists of a collection
of charged particles connected by linear springs in triangular lattice. The
element of elasticity is introduced to balance the repulsive electrostatic force
and to hold the particles together. The elastic triangular lattice also
represents the simplest organization of matter. Charged elastic membranes may be
realized in surfactant bilayer
systems~\cite{hoffmann1994surfactant,oberdisse1996vesicles} and charged lamellar
systems~\cite{deme2002giant}. The geometry and stress of membranes under
various constraints have been extensively
studied~\cite{Ou-Yang1989,capovilla2002lipid,klein2007shaping,tu2013,yong2013elastic,ref1,schweitzer2015model,ou1999geometric,deserno2015fluid}.
The presence of surface charge could significantly influence the elastic
rigidity~\cite{duplantier1990geometrical,ambjornsson2007applying,rowat2004experimental}
and structural
stability~\cite{winterhalter1988effect,andelman1995electrostatic,vernizzi2007faceting,gao2019electrostatic}
of membranes.

In this work, we resort to the combination of numerical simulation and
analytical calculation to determine and analyze the lowest-energy particle
configuration, and to explore the electrostatics-based organization principle of
matter.  We first perform preliminary analytical analysis of the few-body 1D and 2D systems
to demonstrate the complexity of the electrostatic force. The complexity arises
from both the long-range nature of the electrostatic force and its interplay
with the fluctuating geometry of the membrane in the relaxation process. We
further identify the inhomogeneously stretched planar lattice as the
lowest-energy state. Geometric analysis shows the conformal order and the
preserved colinearity and concyclicity in the deformation of the membrane. Based
on these key features, we reveal the Poincar$\rm{\acute{e}}$-Klein mapping
underlying the electrostatics-driven inhomogeneity. The
Poincar$\rm{\acute{e}}$-Klein mapping connects the Poincaré disk and the Klein
disk, which represent two classical models for the hyperbolic plane. We also
discuss the lowest-energy configurations of particles in lattices with patterned
charges of opposite signs for modulating in-plane inhomogeneity and even
creating 3D shapes, which may have connection to metamaterials
design. These results advance our understanding on the organization of matter
under the long-range force from the perspective of geometry.

\section{Model and method}

\begin{figure}[t]
	\centering
	\subfloat[]{\includegraphics[width = .42\linewidth]{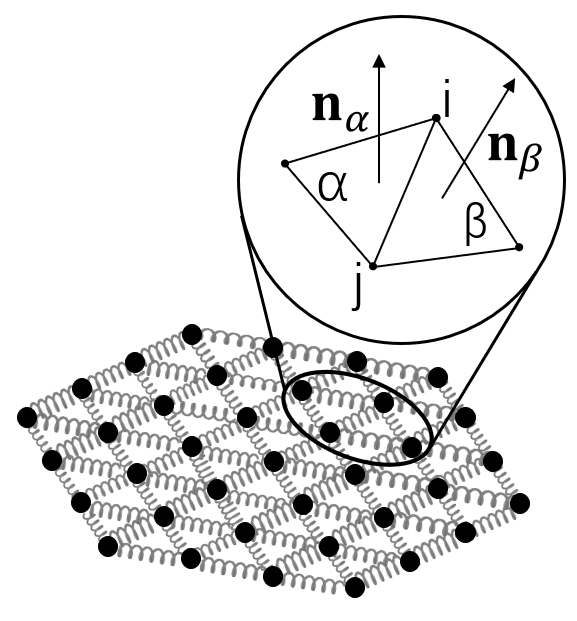}%
		\label{fig1a}}\qquad
	\subfloat[]{\includegraphics[width = .48\linewidth]{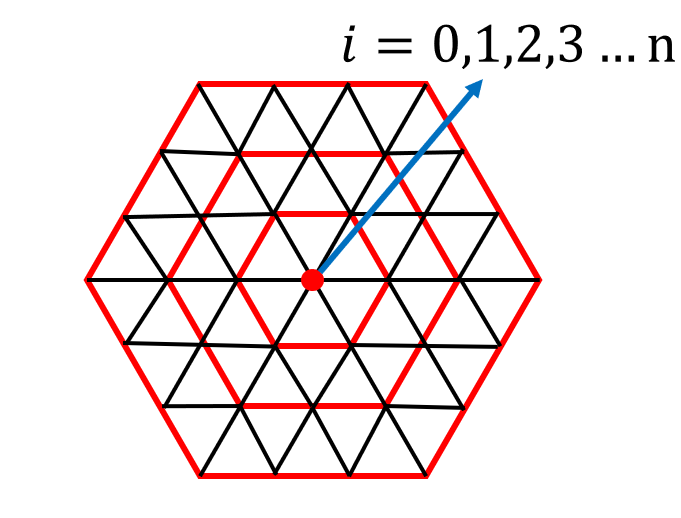}%
		\label{fig1b}}
	\vfill
	\subfloat[]{\includegraphics[width = .45\linewidth]{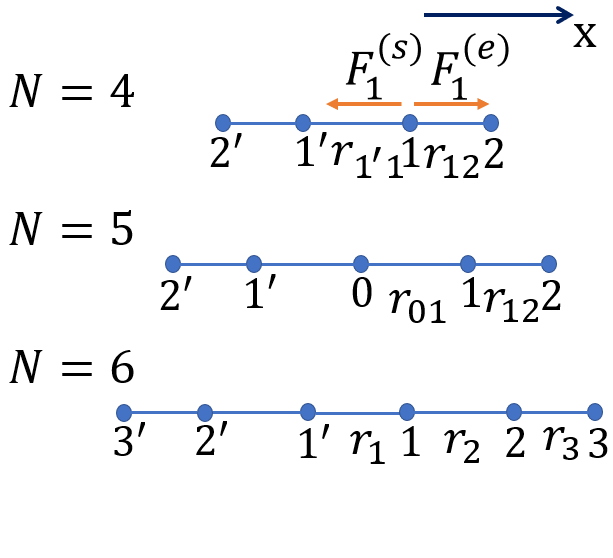}
		\label{fig1c}}\quad
	\subfloat[]{\includegraphics[width = .48\linewidth]{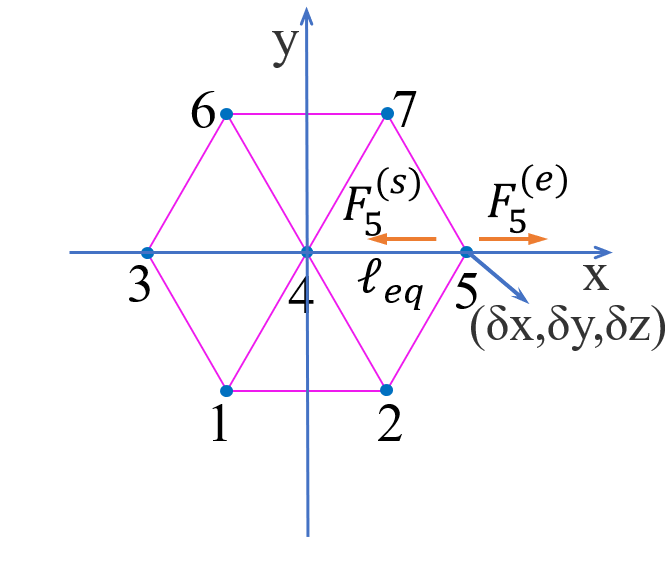}%
		\label{fig1d}}
  \caption{Schematic plots of the model system.  (a) The model consists of a
  collection of charged point particles connected by linear springs in
  triangular lattice of hexagonal shape. (b) Notation for concentric hexagonal
  layers. The central point is denoted as layer 0. The outermost hexagonal
  layer is denoted as layer $n$. (c) Plots of 1D systems for preliminary analysis. 
  (d) Plot of an elementary hexagonal system ($n=1$) for
  analytical analysis. }
	\label{fig1}
\end{figure}

The model consists of a collection of charged point particles in triangular
lattice of hexagonal shape.  Each particle carries a charge of $q_0$. Adjacent
particles are connected by identical linear springs of stiffness $k_s$ and
balance length $\ell_0$. For convenience in discussion, the hexagonal lattice is
divided into a number of layers, as shown in Fig.~\ref{fig1b}. The total number of
particles for a triangular lattice of $n$ layers $N=3n^2+3n+1$. For a given
configuration, the total energy of the system
is~\cite{ref3, ref4}:
\begin{equation}\label{eq_eng}  
	E = k_s\sum_{i\in B} (\ell_i-\ell_0)^2 +
  k_b\sum_{\left\langle\alpha,\beta\right\rangle}
  (1-\mathbf{n}_{\alpha}\cdot\mathbf{n}_{\beta}) + k_e \!\!\!\! \sum_{i,j\in
  V,i\neq j}\!\! \frac{1}{r_{ij}}, 
\end{equation}
where $\ell_i$ is the length of the spring $i$, $\mathbf{n}_{\alpha}$ is the
unit normal vector of the triangle $\alpha$, and $r_{ij}$ is the Euclidean distance of two
particles $i$ and $j$, as shown in Fig.~\ref{fig1a}. The three terms in
Eq.(\ref{eq_eng}) represent the total stretching energy, bending energy and
electrostatic energy, which are defined on the bonds, triangular faces and
points in the triangular lattice, respectively. $k_e=q_0^2 /4\pi \varepsilon_0$,
where $\varepsilon_0$ is the vacuum dielectric constant. In this work, the
units of length and energy are $\ell_0$ and $\epsilon_s$, respectively,
and $\epsilon_s=k_s\ell_0^2$.

We employ the standard annealing Metropolis Monte Carlo algorithm to determine the lowest-energy
states of the model system under the free boundary
condition~\cite{ref6}. The boundary particles are free of any external force except the
elastic and electrostatic force from other particles as defined in Eq.~(\ref{eq_eng}). Given an initial configuration of
tethered particles, we simultaneously move each particle by a random vector
$d \hat{\omega}$, where $\hat{\omega}$ is a random unit vector in 3D space and
$d$ is the step size. The displacement vectors on the particles are statistically
independent. An update of the positions of the particles is denoted as one sweep. The
particle configuration is updated by consecutive sweeps until the system reaches the
lowest-energy state under some specified termination conditions. Specifically, the
relaxation process of the system is terminated if the standard deviation of energy in the
last 5000 sweeps (denoted as $\delta E$) becomes sufficiently small. Typically, $\delta
E$ is as small as $10^{-2}\epsilon_s$. The value of $d$ is in the range of 
$[10^{-4}\ell_0, 10^{-5}\ell_0]$. The value of the final temperature $T_f$ in the last
5000 sweeps is in the range of $[10^{-5}\epsilon_s/k_B, 10^{-6}\epsilon_s/k_B]$, where
$k_B$ is the Boltzmann's constant.

\section{Results and discussion}

This section consists of two subsections.  In Sec.~\ref{sec3a}, we first perform
preliminary analytical analysis of the few-body 1D and 2D systems, and show the
complexity of the long-range electrostatic force.  By numerical simulation, we
identify the inhomogeneously stretched planar lattice as the lowest-energy
state in the parameter space of $(k_b, k_e)$ across several orders of magnitude,
and analyze the inhomogeneity phenomenon in terms of the
bond length. We also discuss the variations of lattice shape and energy in the
relaxation process. In Sec.~\ref{sec3b}, by geometric analysis, we show the
conformal order and the preserved colinearity and concyclicity in the lattice
deformation.  Based these key geometric features, we reveal the
Poincar$\rm{\acute{e}}$-Klein mapping underlying the electrostatics-driven
inhomogeneity, which implies the connection of long-range electrostatic force
and hyperbolic geometry. We also discuss lattices with patterned charges of
opposite signs that may have a connection to metamaterials design, and briefly
discuss possible realizations of the theoretical model.

\subsection{Inhomogeneously stretched planar lattice as the lowest-energy state }\label{sec3a}

We first analyze the analytically tractable 1D few-particle systems, focusing on
the distribution of the particles along the line in mechanical equilibrium. The
schematic plot of the $N$ particles connected by $N-1$ linear springs is shown
in Fig.~\ref{fig1c}. For the case of $N=4$, it can be shown that
$r_{12}<r_{1'1}$, indicating the phenomenon of inhomogeneity under the
long-range repulsive force in a 1D lattice system (see Appendix A for more information).
This conclusion could be generalized to the two boundary springs in 
1D systems of $N>3$ regardless of the values of $k_s$ and $k_e$. For
example, for the cases of $N=5$ and $N=6$, based on the similar argument for the
case of $N=4$, one could derive that $r_{12}<r_{01}$ and $r_3<r_2$,
respectively. By computation, we find that a spring that is closer to the center
of the 1D lattice is subject to stronger pulling force
from the two groups of the particles on its two sides; it is a challenge to
present a rigorous proof due to the long-range nature of the electrostatic
force. In other words, the spring length decreases from the center to the ends
of the 1D lattice in mechanical equilibrium.

For the 2D case, we analyze the mechanically equilibrium
configuration of the elementary single-layer hexagonal system composed of only
seven particles connected by linear springs, as shown in Fig.~\ref{fig1d}.
Combination of analytical and numerical approaches confirms the stability of the
flat hexagonal configuration in Fig.~\ref{fig1d} (see Appendix A for more
information). For a larger hexagonal system, it is expected that similar to the
1D case, the long-range repulsive force may lead to inhomogeneous distribution
of particles. From the perspective of geometry, the inhomogeneity of particle
distribution could be regarded as a modification of the metric structure over
the lattice~\cite{klein2007shaping}.  As such, one may inquire if a larger
hexagonal lattice equipped with a spatially varying metric would be buckled to
the 3D space under the electrostatic force for reducing energy.

To address this question, we employ the standard
annealing Metropolis Monte Carlo algorithm to determine the lowest-energy states
of larger 2D lattice systems~\cite{ref6}. The dimensionless quantity
$\tilde{h}$ is proposed to measure the degree of flatness of the deformed shape:  
\begin{equation}\label{tild_h} 
\tilde{h}=\frac{1}{\langle  d \rangle}\sqrt{\frac{1}{N}\sum_{i=1}^{N} h_i^2},
\end{equation}
where $h_i$ is the distance between particle $i$ and the plane of the reference
triangle spanned by the three particles located at the furthest corners of the
hexagon. The average side length of the reference triangle is denoted as
$\langle  d \rangle$. For the cases of $n=\{1, 2, 3, 4, 5, 6, 7, 8, 9\}$, we
systematically explore the parameter space of $(k_b, k_e)$ across several orders of
magnitude: $k_b=\{0.01, 0.1, 1, 10, 100\}$ and $k_e=\{0.001, 0.01, 0.1, 1, 10,
100\}$. It turns out that the value of $\tilde{h}$ monotonously decreases with
the reduction of temperature, and reaches a small value within the order of
$10^{-3}$ for $T_f \sim 10^{-6}\epsilon_s/k_B$. In the relaxation process, the
value of $\tilde{h}$ shows no dependence on $n$. To conclude, simulations
suggest that the system tends to evolve towards a stretched planar lattice as
the lowest-energy state. To further substantiate this numerical observation, we also relax the lattice
system on the plane (2D relaxation), and compare the energies of the fully
relaxed configurations via 3D and 2D relaxations under a given low temperature,
which are denoted as $E_{\rm{3D}}$ and $E_{\rm{2D}}$, respectively. It is found
that $E_{\rm{2D}}$ is always slightly smaller than $E_{\rm{3D}}$, indicating
that the energy of the fully relaxed lattice via 3D relaxation could be further
lowered by flattening.

\begin{figure}[t]  
	\centering 
	\subfloat[]{\includegraphics[width = .6\linewidth]{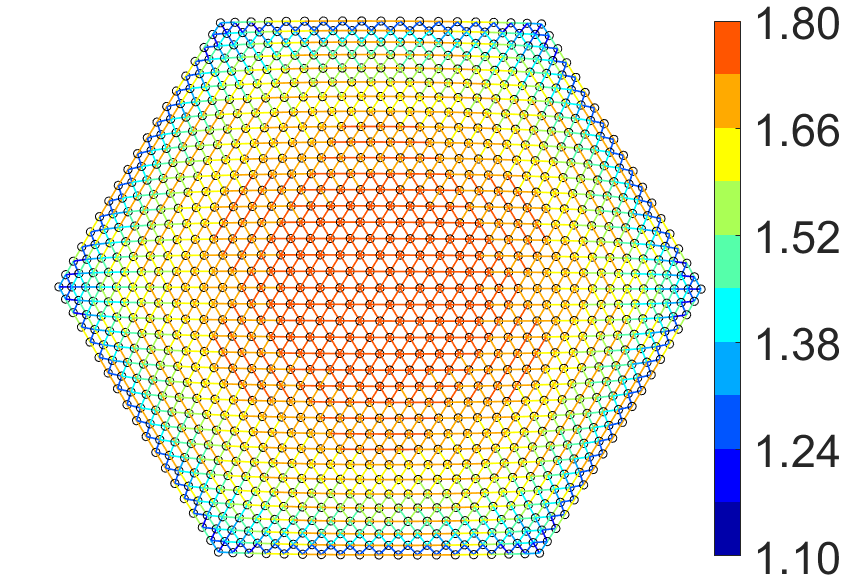}%
		\label{fig4a}}
	\quad
	\subfloat[]{\includegraphics[width = .65\linewidth]{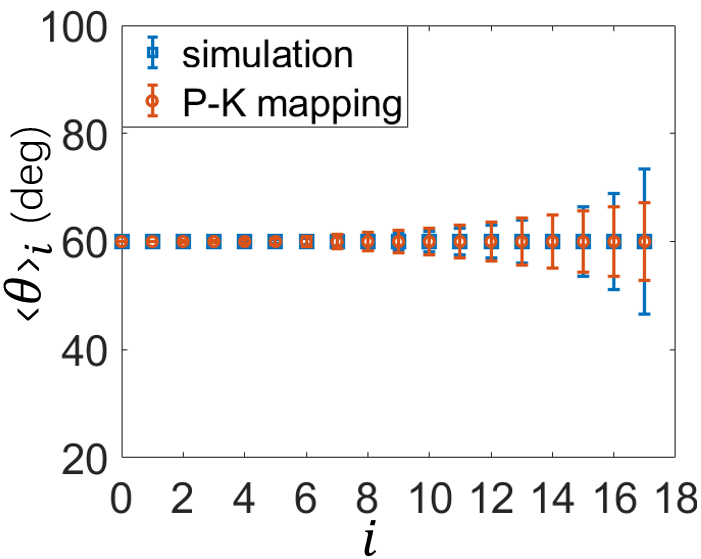}%
		\label{fig4b}}
  \caption{Analysis of bond length and bond angle reveals conformal order in the
  electrostatics-driven inhomogeneity in the lowest-energy configurations.  (a)
  The bond length gradually decreases from about $1.8$ at the center to about
  $1.1$ at the boundary, as indicated by the color legend.  $n=18$. $k_b=0.01$.
  $k_e=0.1$.  (b) Distributions of the average bond angle $\langle\theta\rangle_i$
  in each layer $i$ for both the lowest-energy configuration obtained by
  simulations (blue square) and the configuration generated by the
  Poincar$\rm{\acute{e}}$-Klein mapping (orange circle). The averaging procedure
  in each layer is over all of the bond angles associated with each vertex along
  the layer [see the red hexagons in Fig.~\ref{fig1b}]. The magnitude of the
  standard deviation is indicated by the length of the error bars. The error
  bars associated with simulations are slightly longer than those associated
  with the Poincar$\rm{\acute{e}}$-Klein mapping for $n>14$. $n=18$. $k_b=0.01$.
  $k_e=0.1$. } \label{fig4}
\end{figure}

We also track the variation of the hexagonal
lattice shape in the relaxation process from the perspective of the integral of
the Gaussian curvature for varying values of $n$, $k_e$ and $k_b$: $n=\{1, 2, 3,
4, 5, 6, 7, 8, 9\}$, $k_b=\{0.01, 0.1, 1, 10, 100\}$ and $k_e=\{0.001, 0.01,
0.1, 1, 10, 100\}$. It turns out that the total Gaussian curvature uniformly
converges to zero in all of the cases (see Appendix B for more information). 
We further inquire if the planar lattice as the lowest-energy state is related
to the hexagonal shape. To address this question, we remove $2n_r$ rows of
particles from the originally hexagon-shaped lattice, and perform simulations to
determine the lowest-energy configurations. Simulations also show planar
lattices as the lowest-energy states; the values of $\tilde{h}$ of the
numerically obtained lowest-energy configurations are within the order of
$10^{-3}$. As such, the suppression of
out-of-plane deformation is not unique to the hexagonal shape of the lattice,
and it shall be related to the long-range nature of the electrostatic
force~\cite{yao2013topological}.

\begin{figure}[t]
	\centering
	\includegraphics[width = .65\linewidth]{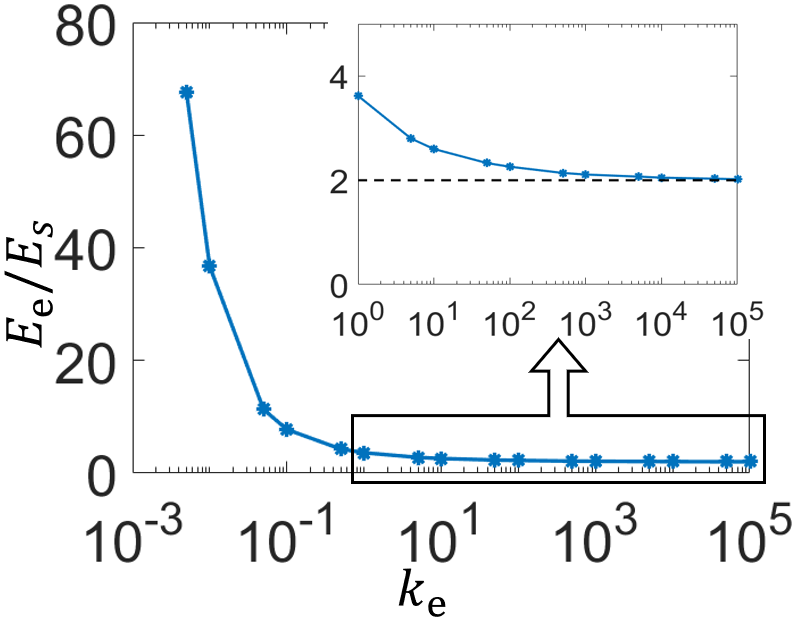}%
  \caption{Convergence of the ratio of the electrostatic energy $E_e$ and the
  stretching energy $E_s$ with the increase of $k_e$. $n=6$. $k_b=0.01$. The
  value of the energy is obtained for fully relaxed configuration. }
  \label{fig3}
\end{figure}

We further analyze the in-plane deformations in the lowest-energy planar
lattices. It is found that the fully relaxed lattice in mechanical equilibrium is
inhomogeneously stretched. The inhomogeneous distribution of particles is
presented in Fig.~\ref{fig4a}. The bond length significantly decreases when
approaching the boundary of the system, implying that the disruption of the
lattice may be initiated from the center of the system.  The degree of
inhomogeneity is controlled by the characteristic length scale $\ell_{es}$.
Simulations at varying values of $k_e$ (and fixed value of $k_s$) show that
increasing $\ell_{es}$ leads to stronger stretching of the springs and larger
variation in the distribution of the bond length over the lattice.

How shall we understand the phenomenon of inhomogeneity in the
electrostatics-driven deformation of the lattice? The crucial element for
shaping the inhomogeneous distribution is the combination of the long-range
nature of the physical interaction and the discrepancy in the ambient
environment of the particles. Specifically, in the hexagonal lattice, each
particle ``sees" distinct distributions of the other particles (except the
particles located at the lattice sites of $C_6$ symmetry), ultimately leading to
spatially varying electrostatic force over the lattice in mechanical
equilibrium. Simulations show that the deformation towards the center of the
hexagonal lattice in mechanical equilibrium is larger for compensating the
stronger electrostatic repulsion.

Regarding the energetics in the relaxation process, the variation of the ratio of the
electrostatic energy $E_e$ to the stretching energy $E_s$ for the stretched lattices in
mechanical equilibrium is plotted in Fig.~\ref{fig3}; the bending energy is vanishingly
small.  With the stretching of the lattice, the ratio $E_e/E_s$ decreases, indicating
the conversion of the electrostatic energy to the stretching energy.  With the increase
of $k_e$ up to $10^5$, the ratio $E_e/E_s$ converges to about 2; statistics of the cases
of varying $n$ from 1 to 9 shows that $E_e/E_s=2.030\pm 0.003$ for $k_e = 10^5$.

To understand the convergence of the ratio $E_e/E_s$ to the common limiting value of
about 2, we present the analytical result for the two-particle system. From the balance
of the electrostatic force and the elastic stretching force, we solve for the equilibrium
configuration of the system and calculate the limiting value of the energy ratio:
\begin{equation} \label{Eratio}
	\lim_{k_e\to\infty}\frac{E_e}{E_s}=\lim_{k_e\to\infty}
  \frac{2\ell_{es}^3}{\ell_{eq}(\ell_{eq} - \ell_0)^2} = 2,
\end{equation}
where the equilibrium distance $\ell_{eq}$ of the two particles could be derived from the
force balance equation and $\ell_{es} = (k_e/k_s)^{1/3}$. The characteristic length
scale $\ell_{es}$ arises from the competition of the electrostatic force and the
elastic force. The expression for $\ell_{eq}$ has the same form as
Eq.(\ref{l_eq}) by replacing $\tilde{\gamma}$ for $\ell_{es}^3/(2\ell_0^3)$. In the limit of
$k_e\to\infty$, $\ell_{eq} \sim k_e^{1/3}$.  The analytical result for the two-particle
system and the numerical results for the large systems suggest that the partition of the
electrostatic energy and the elastic stretching energy conforms to a constant ratio of 2
in the limit of large $k_e$.


\begin{figure*}[t]  
	\centering                                               
	\subfloat[]{\includegraphics[width = .30\linewidth]{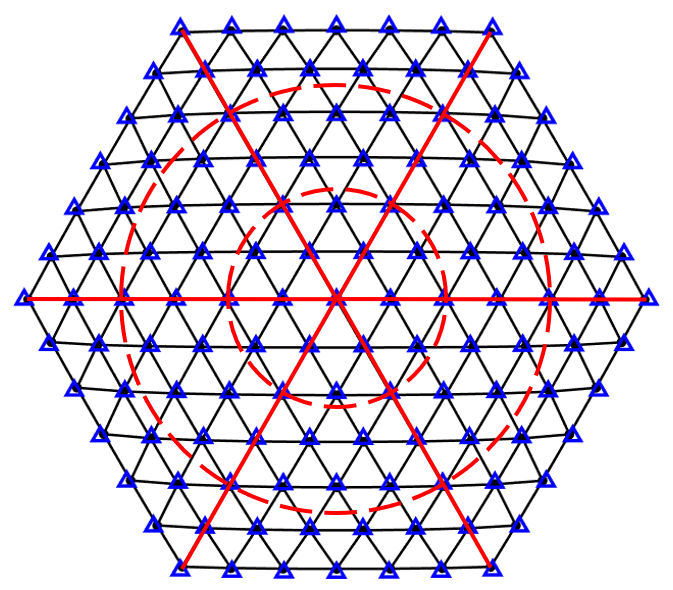}%
		\label{fig5a}} 
	\qquad                                 	
	\subfloat[]{\includegraphics[width = .18\linewidth]{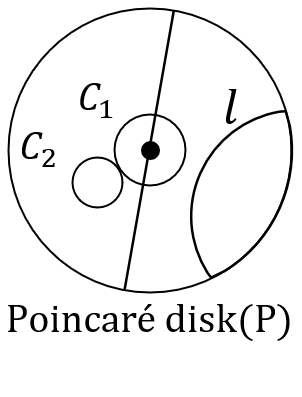}%
		\label{fig5b}}
	\qquad
	\subfloat[]{\includegraphics[width = .18\linewidth]{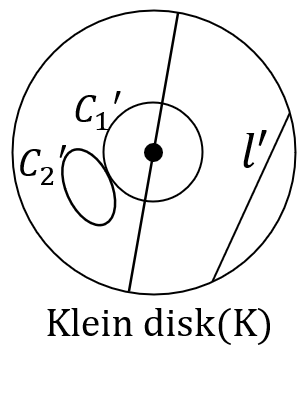}%
		\label{fig5c}}
	\qquad
	\subfloat[]{\includegraphics[width = .22\linewidth]{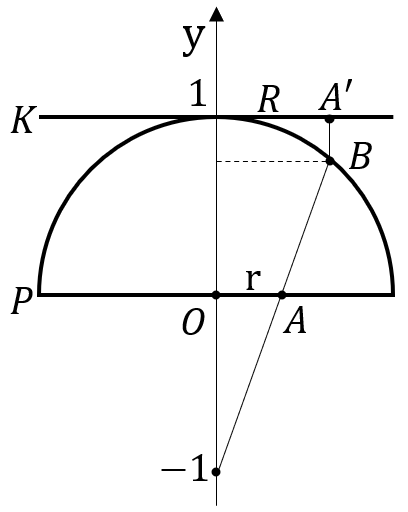}%
		\label{fig5d}}	
  \caption{Poincar$\rm{\acute{e}}$-Klein mapping underlying the
  electrostatics-driven inhomogeneity in the lowest-energy configurations. (a)
  The configuration obtained by simulations (solid black dots) is well fitted by
  the theoretical configuration generated by Poincar$\rm{\acute{e}}$-Klein
  mapping (empty blue triangles). The values of the fitting parameters in
  Eq.(\ref{resca_P-K}) are: $\lambda = 0.032$, $\Gamma = 22.028$. $L= 0.005$.
  $n=6$. $k_b=0.01$. $k_e=0.1$. An infinitely large hyperbolic disk is
  represented in the Poincar$\rm{\acute{e}}$ disk (b) and Klein disk (c). $l$ and
  $l'$ correspond to hyperbolic geodesics in these two models. Circles on the
  hyperbolic plane preserve their shape in the Poincar$\rm{\acute{e}}$ disk
  [$C_1$ and $C_2$ in Fig.~\ref{fig5b}]. In contrast, only concentric circles
  preserve their shape in the Klein disk [$C_1'$ and $C_2'$ in
  Fig.~\ref{fig5c}]. Quantitative information about the two models is
  presented in the paper. (d) Illustration of the
  Poincar$\rm{\acute{e}}$-Klein mapping. The Poincar$\rm{\acute{e}}$ disk and
  the Klein disk are indicated by the two lines labeled as P and K,
  respectively. The semicircle between represents a semi-sphere. Any point in
  the Poincar$\rm{\acute{e}}$ disk (labeled as A) is mapped to $A'$ in the Klein
  disk. The line $BA'$ is perpendicular to the disk $K$.  } \label{fig5}
\end{figure*}

\subsection{Conformal order and Poincar$\rm{\acute{e}}$-Klein mapping underlying the
stretched planar lattice}\label{sec3b}

In the preceding subsection, we have revealed the inhomogeneously stretched
planar lattice as the lowest-energy configuration under the competing
electrostatic force and elastic force. In this subsection, we focus on analyzing
the geometric structures underlying the inhomogeneity phenomenon.

To seek the regularity underlying the inhomogeneous particle distribution, we
analyze the distribution of the bond angle, and inquire if the bond angle is
invariant in the electrostatics-driven deformation of the lattice. In
mathematics, strictly angle-preserved deformation is known as conformal
transformation.  Quasi-conformal order has been revealed in the beautiful
gravity's rainbow formed by electrically charged steel beads in gravity
field~\cite{pieranski1989gravity,rothen1993conformal,wojciechowski1996minimum}.
Recently, quasi-conformal order has been reported in the inhomogeneous
packings of long-range repulsive particles confined on the
disk~\cite{soni2018emergent}, self-assembled vortices in a type-II
superconducting disk~\cite{vortexSilva2019} and Lennard-Jones particles confined
on the sphere~\cite{chen2022geometry}.

The plot of the bond angle along the layers of the lattice is shown in
Fig.~\ref{fig4b}. A key observation is that the average bond angle is
uniformly $60$ degrees; the standard deviation increases when approaching the
boundary.  Therefore, the deformation of the interior region of the lattice
could be regarded as conformal. In other words, each triangular cell in the
interior region experiences isotropic expansion according to the geometric
property of conformal transformation~\cite{pieranski1989gravity}. Furthermore,
in the sense of average bond angle, the lattice deformation is
quasi-conformal.  We also vary the values of $n$ and $k_e$ (see Appendix B).
It is found that the average bond angle is uniformly $60$ degrees as the value
of $k_e$ is varied by several orders of magnitude, indicating that the
preserved average bond angle is a common feature in the electrostatics-driven
deformation of the lattice.  Note that in the measurement of bond angles in the
deformed lattice, the vertices are connected by straight lines. However, the
lattice is bent in the continuum limit. Using smooth curves instead of straight
lines to connect vertices may reduce the magnitude of the error bar especially
near the boundary in Fig.~\ref{fig4b}~\cite{pieranski1989gravity}.

In addition to the invariant bond angle, are there any other invariant properties in the
deformation of the lattice? Scrutiny of the lowest-energy configurations reveals
some common features that are illustrated in Fig.~\ref{fig5a}. First, the
particles remain on the principal axes (the three solid red lines) of the triangular
latticed in the deformation. Second, the particles remain on the concentric
circles (the dashed red circle) in the deformation. For the case in
Fig.~\ref{fig5a}, the mean deviations of the points both from the three principal
axes and from the concentric circles are at the order of $10^{-2}$. Third, the
bond length is appreciably shorter when approaching the boundary of the system.
These geometric features inspire us to explore the mathematical structure
underlying the deformation. Specifically, we search for a mapping that could
reproduce these key features in the in-plane deformation.

The Poincar$\rm{\acute{e}}$ disk and the Klein disk represent two kinds of
important models for representing the infinitely large 2D hyperbolic plane in an
open unit disk in 2D Euclidean space~\cite{benedetti2012lectures}. Notably, the
concept of hyperbolic geometry is widely applied to the analysis of complex
networks in recent years~\cite{krioukov2010hyperbolic,BATTISTON20201}.
Hyperbolic geodesics are mapped to arcs in the Poincar$\rm{\acute{e}}$ disk, and
to straight lines in the Klein disk, as indicated by $l$ and $l'$ in
Fig.~\ref{fig5b} and Fig.~\ref{fig5c}.  Circles on the hyperbolic plane preserve
their shape in the Poincar$\rm{\acute{e}}$ disk [$C_1$ and $C_2$ in
Fig.~\ref{fig5b}]. In contrast, only concentric circles preserve their shape in
the Klein disk [$C_1'$ and $C_2'$ in Fig.~\ref{fig5c}]. These features could be
rigorously derived from the metric tensor over the disk. The line element $ds$
in the Poincar$\rm{\acute{e}}$ disk is~\cite{benedetti2012lectures}
\begin{equation}\label{Pmetric}
	ds^2 = \frac{4\Vert \mathbf{dx}\Vert^2}{(1-\Vert \mathbf{x}\Vert^2)^2},
\end{equation}
where $\mathbf{x}=(x_1, x_2)$, and $\Vert\cdot\Vert$ denotes the Euclidean norm.
$\Vert\mathbf{x}\Vert\leqslant 1$. 
The line element $ds$ in the Klein disk is~\cite{benedetti2012lectures}
\begin{equation}\label{Kmetric}
	ds^2 = \frac{\Vert \mathbf{dx}\Vert^2}{1-\Vert \mathbf{x}\Vert^2} +
  \frac{(\mathbf{x}\cdot\mathbf{dx})^2}{(1-\Vert \mathbf{x}\Vert^2)^2},
\end{equation}	
where $\Vert\mathbf{x}\Vert\leqslant 1$.

These two kinds of models are connected by the Poincar$\rm{\acute{e}}$-Klein
mapping~\cite{benedetti2012lectures} (see Appendix D for more information): 
\begin{equation}\label{P-K}
	f(z)=\frac{2z}{1+|z|^2},\quad |z|\leqslant 1
\end{equation} 
Equation~(\ref{P-K}) states that a point $A$ at $z$ in the Poincar$\rm{\acute{e}}$ disk is mapped
to the point $A'$ at $f(z)$ in the Klein disk. The Poincar$\rm{\acute{e}}$-Klein mapping in
Eq.~(\ref{P-K}) is illustrated in Fig.~\ref{fig5d}. The semicircle between the lines
K and P represents a hemisphere of unit radius. The Poincar$\rm{\acute{e}}$ disk and the Klein disk are
represented by the lines P and K, respectively.

In order to characterize the deformation of the lattice, we rescale
Eq.(\ref{P-K}) by introducing two parameters $\lambda$ and $\Gamma$:
\begin{equation}\label{resca_P-K}
	w(z; \lambda, \Gamma)=\Gamma\frac{2\lambda z}{1+\lambda^2 |z|^2}.
\end{equation}
The generalized Poincar$\rm{\acute{e}}$-Klein mapping in Eq.~(\ref{resca_P-K})
preserves the key features in the deformation of the lattice. First, two
arbitrary points on a radial line in the $z$ plane denoted as $z_1=r_1
e^{i\theta}$ and $z_2=r_2 e^{i\theta}$ are mapped to $w_1=\alpha_1 e^{i\theta}$
and $w_2= \alpha_2 e^{i\theta}$, where $\alpha_i = \Gamma\frac{2\lambda
r_i}{1+\lambda^2 r_i^2}$. The new points are still on the same radial line.
Second, two arbitrary points on a concentric circle denoted as
$z_1=re^{i\theta_1}$ and $z_2=re^{i\theta_2}$ are mapped to $w_1=\beta
e^{i\theta_1}$ and $w_2= \beta e^{i\theta_2}$, where $\beta=
\Gamma\frac{2\lambda r}{1+\lambda^2 r^2}$. The new points remain on a concentric
circle.

Now, we search for the optimal values of the parameters $\lambda$ and $\Gamma$ in
Eq.(\ref{resca_P-K}) for fitting the deformation of the lattice. Specifically, the optimal
values of $\lambda$ and $\Gamma$ are determined by minimizing the deviation of the 
configuration generated by the Poincar$\rm{\acute{e}}$-Klein mapping from the lowest-energy
configuration obtained in simulations. The dimensionless quantity to be minimized is:
$L=\langle\delta r\rangle/\langle\ell \rangle$. $\langle\delta
r\rangle=\sqrt{\sum_{i=1}^{N} [(x^{s}_i-x^{m}_i)^2+(y^{s}_i-y^{m}_i)^2]}/N$, where the
superscripts of s and m indicate the data from the simulation and the Poincar$\rm{\acute{e}}$-Klein
mapping. $\langle\ell \rangle$ is the average bond length in the lowest-energy
configurations. We work in the discretized parameter space of $\lambda$ and
$\Gamma$. $\lambda \in (0, 1)$ and $\Gamma \in (0,100)$ in the resolution of
$\delta\lambda=\delta\Gamma=10^{-3}$.

Figure~\ref{fig5a} shows that the configuration generated by the
Poincar$\rm{\acute{e}}$-Klein mapping (marked as empty blue triangles) agrees
well with the deformed lattice in mechanical equilibrium (solid black dots
connected by bonds) for the case of $n=6$. We also check larger systems up to
$n=18$ and a series of values of $k_e$ ranging from $0.001$ to $100$, and find good agreement of
the configurations generated by the Poincar$\rm{\acute{e}}$-Klein mapping and
those by simulations. For example, for $k_e=100$, the values of $L$ (an
indicator for the deviation of the lattices generated by simulations and by the
Poincar$\rm{\acute{e}}$-Klein mapping) for both cases of $n=6$ and $n=18$ are
approximately equal to 0.006; the lattices are strongly stretched for $k_e=100$.

\begin{figure}[t]  
	\centering 
	\includegraphics[width =1\linewidth]{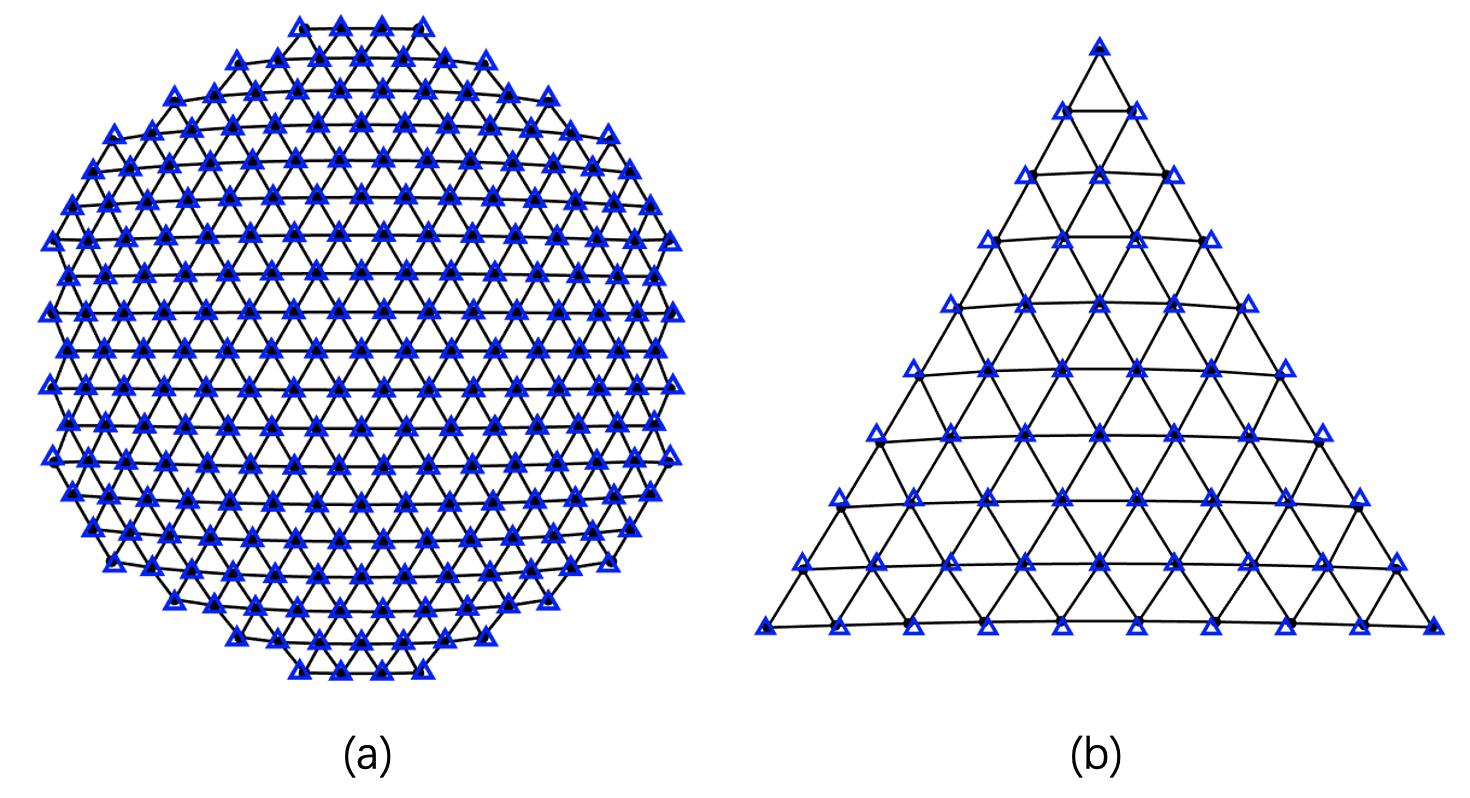}%
  \caption{Presence of Poincar$\rm{\acute{e}}$-Klein mapping in
  nonhexagonal lowest-energy lattices. The lowest-energy configurations of the
  circular (a) and triangular (b) lattices obtained by simulations (solid black
  dots) are well fitted by the theoretical configurations generated by
  Poincar$\rm{\acute{e}}$-Klein mapping (empty blue triangles). The values of the fitting parameters in
  Eq.(\ref{resca_P-K}) are: (a) $\lambda=0.04$, $\Gamma=19.20$. $L\approx
  0.003$.  (b)  $\lambda=0.01$, $\Gamma=63.88$. $L\approx 0.010$. 
  In both cases, $k_b=0.01$, $k_e=0.1$. }
	\label{nonhex_add}
\end{figure}

Furthermore, we check nonhexagonal cases. For the circular and the
triangular lattices as shown in Fig.~\ref{nonhex_add}, simulations show that the
lowest-energy configurations of both cases are still well fitted by the
Poincar$\rm{\acute{e}}$-Klein mapping. Note that due to the
discreteness of the crystalline lattice, the circular lattice in
Fig.~\ref{nonhex_add}(a) possesses the $C_6$ symmetry. Recall that the
existence of the Poincar$\rm{\acute{e}}$-Klein mapping relies on the preserved
colinearity and concyclicity in lattice deformation. Therefore, it is expected
that the Poincar$\rm{\acute{e}}$-Klein mapping structure exists in lattices of
certain symmetries for which the key features of colinearity and concyclicity
are preserved in the deformation. To study the deviation of the deformations with
respect to the Poincar$\rm{\acute{e}}$-Klein mapping, we also check truncated hexagonal
shapes. Such anisotropic shapes are created by removing $n_r$ rows of
particles from two opposite edges in the hexagon-shaped lattice (see Appendix B
for the shapes of truncated hexagonal membranes). In the
cases of $n_r=\{0, 2, 4, 8\}$ for $n=9$, $k_b=0.01$ and $k_e=1$, we analyze the 
deviation of the lowest-energy lattices generated by simulations and by the
Poincar$\rm{\acute{e}}$-Klein mapping, which is characterized by $L$. It is
found that the value of $L$ monotonously increases from $0.004$, $0.017$,
$0.031$ to $0.042$ as the value of $n_r$ is increased from $0$, $2$, $4$ to $8$,
respectively. In other words, the deviation of the deformation with respect to
the Poincar$\rm{\acute{e}}$-Klein mapping is enlarged with the enhanced
anisotropy of the lattice shape.

\begin{figure*}[t]  
	\centering 
	\includegraphics[width =1\linewidth]{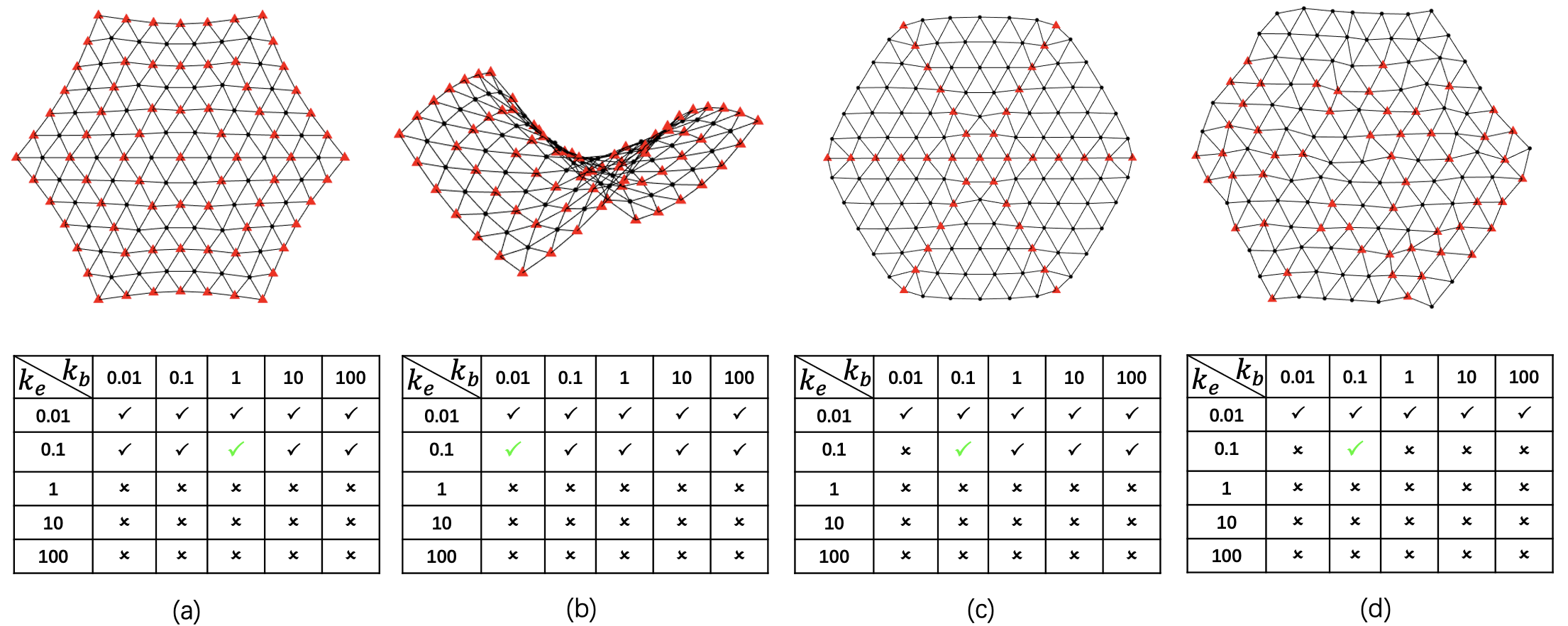}%
  \caption{Lowest-energy states of the lattices under typical distributions of
  charges. The red triangles represent particles carrying charges of the same sign;
  the remaining particles are oppositely charged. The cases that charges of
  opposite signs stick together are marked by the cross symbols in the tables
  for the corresponding charge distributions; nonstick cases are marked by the
  tick symbols. The values of $k_e$ and $k_b$ for the lattice shapes
  are in green in the tables. $\{k_e, k_b\} = \{0.1, 1\}$
  (a), $\{0.1, 0.01\}$ (b), $\{0.1, 0.1\}$ (c) and $\{0.1, 0.1\}$ (d). $k_s=1$. }
	\label{charge_add}
\end{figure*}

In Fig.~\ref{fig4b}, we also show the distribution of the bond angle in each layer
of the lattice for the configuration generated by the Poincar$\rm{\acute{e}}$-Klein mapping. While
the deformation in the central region of the lattice could be regarded as
conformal, the lattice deformation as a whole is quasiconformal in the sense of
average bond angle. In fact, the condition for the formation of strictly
conformal lattice is critical; no uniform field could stabilize a perfect
conformal crystal with the inverse power
potential~\cite{wojciechowski1996minimum}. Strictly speaking, for the
deformation defined by Eq.(\ref{resca_P-K}), only the deformation at the origin
is conformal for the following reason.

According to the geometric property of conformal transformation, no shearing
occurs in the deformation for preserving the angle. For the deformation
characterized by $g(z)=u(x,y)+iv(x,y)$ ($z=x+iy$), the strain field could be
derived as~\cite{rothen1993conformal}
\begin{eqnarray}\label{uxy}
  u_{xx} &=&   \frac{1}{2}\left[\left( \frac{\partial u}{\partial
  x}\right )^2 + \left(\frac{\partial v}{\partial
  x} \right)^2 -1 \right], \nonumber \\
  u_{yy} &=&  \frac{1}{2}\left[\left( \frac{\partial u}{\partial
  y}\right )^2 + \left(\frac{\partial v}{\partial
  y} \right)^2 -1 \right], \nonumber \\ 
  u_{xy} &=&  u_{yx} = \frac{1}{2}\left( \frac{\partial u}{\partial
  x}\frac{\partial u}{\partial y} + \frac{\partial v}{\partial
  x}\frac{\partial v}{\partial y}  \right).
\end{eqnarray}
Note that a typo in Ref.~\cite{rothen1993conformal} has been corrected in
Eqs.(\ref{uxy}).
It is straightforward to check that $u_{xy}=u_{yx}=0$ and $u_{xx}=u_{yy}$ if $g(z)$
satisfies the Cauchy-Riemann conditions~\cite{churchill2014}:
\begin{equation}\label{eqCR}
  \frac{\partial u}{\partial x}=\frac{\partial v}{\partial y},\quad
  \frac{\partial u}{\partial y}=-\frac{\partial v}{\partial x}.
\end{equation}
For $w(z; \lambda, \Gamma)$ in Eq.(\ref{resca_P-K}), the Cauchy-Riemann conditions
are satisfied only at $z=0$.  As such, the deformation as characterized by
Eq.(\ref{resca_P-K}) is strictly conformal only at the origin of the system. The
$z=0$ point in the continuum description of Eq.(\ref{resca_P-K}) corresponds to an
area containing a number of particles in the discretized particle system.

Here, we emphasize that the revealed Poincar$\rm{\acute{e}}$-Klein mapping
underlying the electrostatics-driven inhomogeneity in tethered membranes is
based on the invariant properties in the large deformation of the lattice.  Note
that, considering the complexity of the long-range interaction in organizing
particles, it is important to focus on the invariant elements in the
deformation. Specifically, the key features of the preserved colinearity and
concyclicity in the strongly deformed lattice provide important clues for
uncovering the connection with hyperbolic geometry. In comparison with the
classical scenario of a stretched elastic disk by applying force on the
boundary, which results in homogeneous strain field~\cite{Landau1986} (see
Appendix C for more information), the electrostatic force creates inhomogeneity
in the hexagon-shaped lattice system, which in turn encodes the information of how
electrostatics regulates the distribution of matter. As such, geometric analysis
of the inhomogeneity phenomenon represents an important approach for extracting
such information. For example, in the context of this work, the unique geometric
perspective of the Poincaré-Klein mapping is useful for capturing the
characteristic invariant features of colinearity and concyclicity in the lattice
deformation as well as suggesting the connection of long-range force and
hyperbolic geometry.

We also discuss the extension of the current study by tailoring the
distributions of charges of different signs for targeted shapes with potential
connection to metamaterials design, as inspired by the suggestion of the
anonymous referee. The lowest-energy states of the lattices under typical
distributions of charges are presented in Fig.~\ref{charge_add}, where the red
triangles represent particles carrying charges of the same sign; the remaining
particles are oppositely charged. In Figs.~\ref{charge_add}(a) and
\ref{charge_add}(b), the red triangles are along
concentric hexagons. In Figs.~\ref{charge_add}(c) and \ref{charge_add}(d), the
red triangles are along the three principal axes and are randomly distributed,
respectively. These systems are relaxed by the annealing Metropolis Monte Carlo
algorithm. The expression for the total energy of the system is the same as
Eq.~(\ref{eq_eng}) except the last term whose sign becomes negative if the pair
of points $i$ and $j$ carry opposite charges. Systematic study of the lattice
system at varying values of $k_e$ and $k_b$ shows that charges of opposite signs
stick together for large $k_e$. Such cases are indicated by the cross symbols in
the tables for the corresponding charge distributions in Fig.~\ref{charge_add};
nonstick cases are marked by the tick symbols. The values of $k_e$ and $k_b$
for the presented lattice shapes are in green in the tables.

From Figs.~\ref{charge_add}(a) and \ref{charge_add}(b), we see that the originally
quasiplanar lattice becomes a saddle-like shape as the bending rigidity is
reduced from $k_b=1$ to $k_b=0.01$; the amount of the out-of-plane deformation in
the lattice in Fig.~\ref{charge_add}(a) is at the order of $10^{-2}$. The
saddle-like shape becomes more curved with the reduction of the value of
$k_b$ from 0.1 to 0.01. The formation of the negatively curved saddle-like shape could be
attributed to the circumferential expansion and the radial shrinking of the
lattice under the electrostatic force~\cite{benedetti2012lectures}.
The lowest-energy shapes of the lattices in Figs.~\ref{charge_add}(c) and
\ref{charge_add}(d) are quasiplanar. The amount of the out-of-plane
deformations is at the order of $10^{-2}$. These results demonstrate the
modulation of both the lattice shape and the in-plane particle arrangement by
the design of charge pattern.

We finally briefly discuss possible experimental realizations of the theoretical
model. One may employ charged beads to fabricate the experimental system
as introduced in a recent work on the experimental realization of the theoretical
model of charged beads on a string in polymer science~\cite{charged09}. In
Ref.~\cite{charged09}, the system comprises electrically charged,
millimeter-scale Teflon and Nylon beads placed on a paper over a flat aluminum
sheet. To fabricate a tethered membrane corresponding to our theoretical
model, one may employ these electrically charged beads and introduce elastic
springs. In addition to the above-mentioned macroscopic beads system, one may
also resort to electrically charged colloids confined on liquid interface for
constructing a flat tethered membrane. The effective interactions between
colloids are tunable. To obtain the approximately Coulombic interaction, the
Debye length in the Yukawa interaction could be increased by reducing the salt
concentration~\cite{Walker2011,kumar2019} and minimizing the image-charge
effect~\cite{PhysRevE.92.062306}.

\section{Conclusion}

In summary, the theme of this work is to analyze the deformations of tethered
membranes under the combined electrostatic and elastic forces. The triangular
lattice system provides a suitable model for understanding the organization of
matter by the long-range electrostatic force. The long-range nature of the
electrostatic force and its complicated interplay with the fluctuating geometry
of the membrane impose a challenge to this problem. By combination of numerical
simulation and analytical geometric analysis, we show the crucial role of the
long-range electrostatic force for the suppression of out-of-plane deformations
and the formation of inhomogeneity in the lowest-energy configurations.
Especially, we highlight the revealed Poincar$\rm{\acute{e}}$-Klein mapping that
captures the invariant geometric features in the inhomogeneous deformation and
also implies the connection of long-range repulsive force and hyperbolic
geometry. This work suggests the geometric analysis as a promising approach for
elucidating the inhomogeneous organization of matter under the long-range force.
As inspired by the suggestion of the anonymous referee, it is of interest to
extend the current study by tailoring the charge patterns of different signs for
targeted shapes and particle distributions, which may have connection to
metamaterials design. Preliminary results demonstrate the modulation of in-plane
inhomogeneity and the appearance of saddle-like shapes under different kinds of
charge patterns.

\section*{Appendix A: Preliminary analytical analysis of 1D \& 2D systems}

The schematic plot of 1D system consisting of $N$ particles connected by $N-1$
linear springs is shown in Fig.~\ref{fig1c}. For the trivial case of $N=3$, the
lengths of the two springs are identical. For the case of $N=4$,
$r_{12}<r_{1'1}$. A proof by $reductio\ ad\ absurdum$ is presented below. Suppose
$r_{12}\geqslant r_{1'1}$, the total elastic force on particle 1,
$\vec{F}^{(s)}_1=k_s(r_{12}-r_{1'1})\hat{x}$, is either pointing rightward or
zero. Consequently, the total electrostatic force $\vec{F}^{(e)}_1$ on particle
1 is either pointing leftward or zero to satisfy the force balance condition.
However, the supposed condition of $r_{12}\geqslant r_{1'1}$ leads to an
electrostatic force on particle 1 pointing rightward. This contradiction
indicates the impossibility of $r_{12}\geqslant r_{1'1}$. Therefore, for the
case of $N=4$, $r_{12}$ is always smaller than $r_{1'1}$.

\begin{figure*}[th]
	\centering
	\includegraphics[width = 0.85\linewidth]{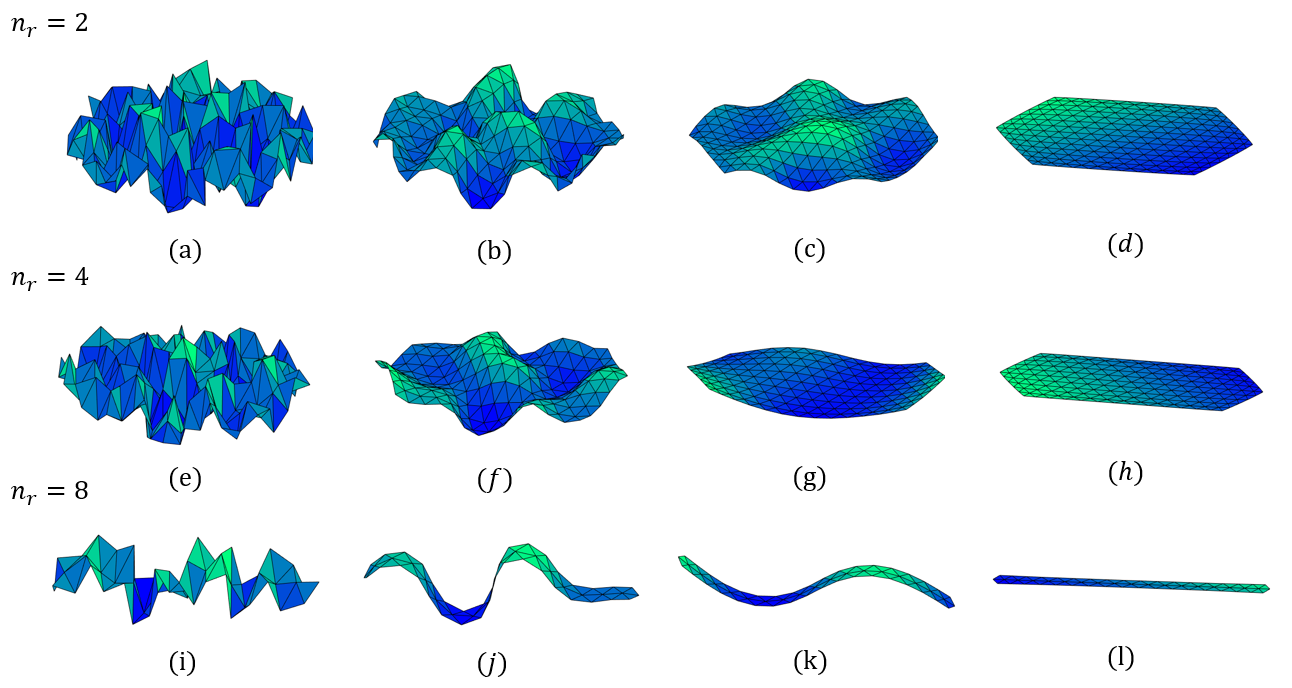}
  \caption{Relaxations of truncated hexagonal membranes under the electrostatic
  force. $n_r$ rows of particles are removed from two opposite edges in the
  hexagon-shaped lattice. Typical membrane shapes in the relaxation process are
  presented for each case. For visual convenience, the color of the figures is
  rendered by the heights of the triangles in the lattice; the brighter
  triangles are located at a larger height. In the lowest-energy
  shapes, $\tilde{h}=8.02\times10^{-5}$(d), $9.78\times10^{-5}$(h), and
  $2.53\times10^{-4}$(l), respectively. $n=9$, $k_b=0.01$, $k_e=100$.  }
  \label{fig4.1}
\end{figure*}

For the 2D case, we perform analysis of the mechanically equilibrium
configuration of the elementary single-layer hexagonal system composed of only
seven particles connected by linear springs, as shown in Fig.~\ref{fig1d}. The
hexagonal configuration is confined on the plane. For an arbitrary particle on
the boundary denoted as 5 in Fig.~\ref{fig1d}, the balance of the electrostatic
force $F_5^{(e)}$ and the elastic stretching force $F_5^{(s)}$ along the $x$ axis
leads to
\begin{equation} \label{fc_balan}
	2k_s(\ell_{eq}-\ell_0)=\frac{k_e}{\ell_{eq}^2}(\frac{\sqrt{3}}{3} +
  \frac{9}{4}), 
\end{equation}
where $\ell_{eq}$ is the equilibrium bond length, and 
\begin{equation} \label{ell_es}
  \ell_{es} = \left(\frac{k_e}{k_s}\right)^{\frac{1}{3}}. 
\end{equation}
From Eq.~(\ref{fc_balan}), we obtain
\begin{equation} \label{l_eq}
	\begin{aligned}
    \frac{\ell_{eq}}{\ell_0} =&  \left(
    \tilde{\gamma}+\frac{1}{27}+\sqrt{(\tilde{\gamma}+\frac{1}{27})^2-(\frac{1}{27})^2}
    \right) ^{\frac{1}{3}} \\
    &+\left(\tilde{\gamma}+\frac{1}{27}-\sqrt{(\tilde{\gamma}+\frac{1}{27})^2-(\frac{1}{27})^2}\right)^{\frac{1}{3}}+\frac{1}{3}, \\
	\end{aligned}
\end{equation}	
where
\begin{eqnarray}
\tilde{\gamma}=(\frac{\sqrt{3}}{12} +
\frac{9}{16})\frac{\ell_{es}^3}{\ell_0^3}.
\end{eqnarray}
The dimensionless quantity $\tilde{\gamma}$ reflects the competition of the
electrostatic force and the elastic force.

\begin{figure}[th]
	\centering
	\subfloat[]{\includegraphics[width = .5\linewidth]{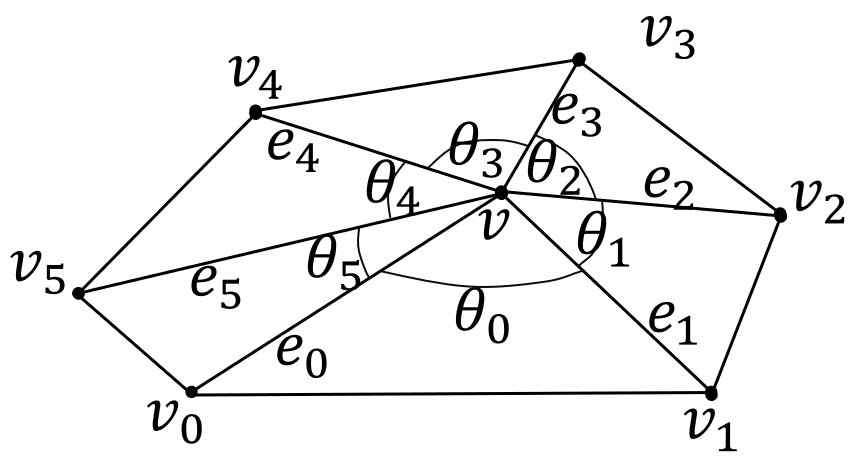}%
		\label{fig2a}}
	\vfill
	\subfloat[]{\includegraphics[width = .6\linewidth]{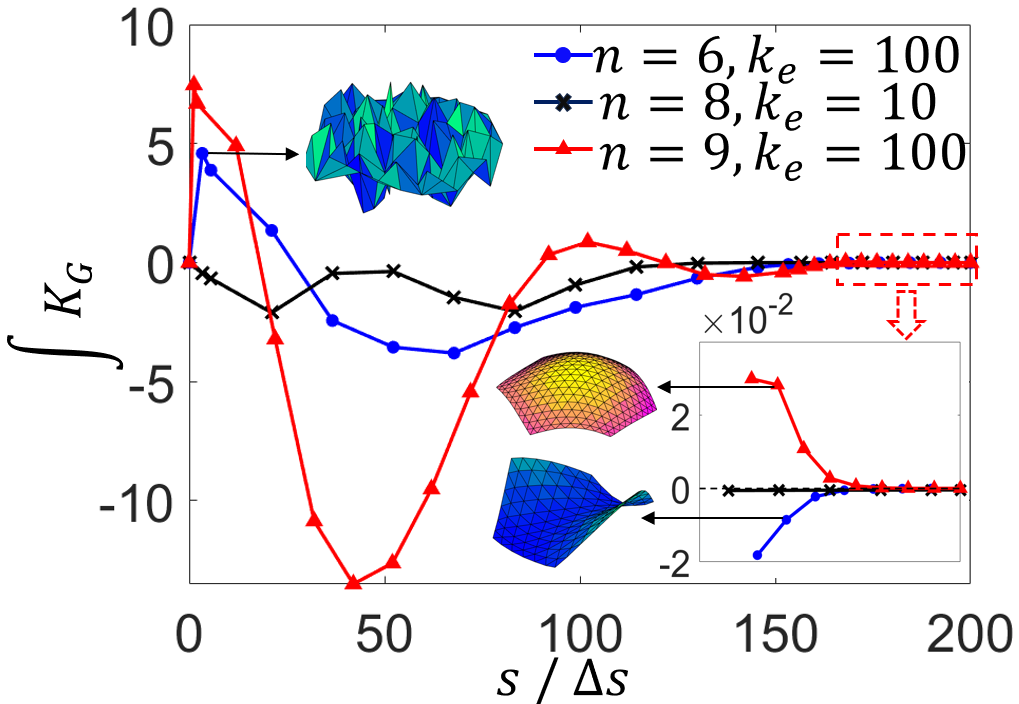}%
		\label{fig2b}}
  \caption{Analysis of the relaxation process of the 2D system from the
  perspective of the variation of the Gaussian curvature. (a) Notations for the calculation of the
  Gaussian curvature in the main text. (b) Variation of the total Gaussian
  curvature over the whole lattice in the relaxation process. $s$ represents the number of
  simulation steps. $\Delta s =9000,9000,25000$ for the curves of $n=6$ $n=8$
  and $n=9$, respectively. $k_b=0.01$. The zoomed-in inset shows the convergence
  of the integrated Gaussian curvature to zero. }
	\label{fig2}
\end{figure}

It is of interest to discuss Eq.(\ref{l_eq}) in the limiting cases of
$\tilde{\gamma}\to 0$ (strong spring stiffness) and $\tilde{\gamma}\to \infty$
(strong electrostatic effect). As $\tilde{\gamma}\to 0$, 
\begin{equation} \label{l_eq_gamma0}
  \frac{\ell_{eq}}{\ell_0} = 1+2\tilde{\gamma}-8\tilde{\gamma}^2+o(\tilde{\gamma}^2).
\end{equation}
It is reasonable that $\ell_{eq}=\ell_0$ when $\tilde{\gamma}=0$ (i.e., $k_e=0$).
In contrast, the asymptotic expression for $\ell_{eq}$ as $\tilde{\gamma}\to
\infty$ is
\begin{equation}\label{l_eq_gamma_inf}
  \frac{\ell_{eq}}{\ell_0} = (2\tilde{\gamma})^\frac{1}{3} = \beta
  \frac{\ell_{es}}{\ell_0},
\end{equation}
where $\beta=(\frac{\sqrt{3}}{6} + \frac{9}{8})^\frac{1}{3}\approx 1.12$.  The exponent
 $1/3$ in Eq.~(\ref{l_eq_gamma_inf}) originates from the cubic term in Eq.~(\ref{fc_balan}),
which arises from the competition of the elastic force and the electrostatic force.

Previous study shows the counterintuitive instability of charged elastic rings
under the long-range electrostatic force~\cite{yao2023ring}. Here, it is of interest to
examine the possible out-of-plane deformation of the elementary hexagonal
configuration in Fig.~\ref{fig1d}. To this end, we first impose a random
perturbation $\delta \vec{r}$ in 3D space on an arbitrary particle labeled 5,
and calculate the variation of energy. $\delta \vec{r}=(\delta x,\delta y,\delta
z)$. The resulting variation of the total energy is 
\begin{equation} \label{dE}
	\begin{aligned}
		\delta E=& k_e\sum_{i=1,i\neq 5}^{7}(\frac{1}{|\vec{r}_{i5}+\delta \vec{r}|}-\frac{1}{|\vec{r}_{i5}|}) \\
		&+\frac{k_s}{2}\sum_{i=2,4,7}[(|\vec{r}_{i5}+\delta \vec{r}|-\ell_0)^2-(|\vec{r}_{i5}|-\ell_0)^2], \\
	\end{aligned}
\end{equation} 	
where $|\vec{r}_{i5}|$ is the distance between particle $i$ and 5. Keeping up to the second order terms,
Eq.(\ref{dE}) becomes
\begin{equation} \label{}
	\begin{aligned}
        \delta E= C_{11}\delta x^2 + C_{22}\delta y^2 + C_{33}\delta z^2, \\
	\end{aligned}		
\end{equation}
where 
\begin{equation} \label{}
	\begin{aligned}
    &\frac{C_{11}}{k_s}=\frac{\ell_{es}^3}{4\ell_{eq}^3}(\frac{19\sqrt{3}}{18}+\frac{55}{8})+\frac{3}{4}, \\
    &\frac{C_{22}}{k_s}=\frac{\ell_{es}^3}{4\ell_{eq}^3}(\frac{7\sqrt{3}}{18}+\frac{49}{8})+\frac{3}{4}, \\
    &\frac{C_{33}}{k_s}=\frac{\ell_{es}^3}{4\ell_{eq}^3}(\frac{5\sqrt{3}}{9}+\frac{1}{2}). \\
	\end{aligned}	
\end{equation} 
All of these coefficients are positive. Therefore, $\delta E > 0$. In other words, the system is stable under arbitrary
perturbation of any particle.

We further examine the case where all of the particles in Fig.~\ref{fig1d} are
perturbed by performing numerical simulations. The magnitude of the
statistically independent random perturbation on each particle is restricted
within the range from $-0.3\ell_{eq}$ to $0.3\ell_{eq}$. For each given value
of $k_e$ in $\{0.001,0.01,0.1,1,10,100\}$, we generate $10^4$ statistically
independent particle configurations, and compute the variation of energy by
Eq.~(\ref{dE}). It turns out that the energy is always increased upon simultaneous
perturbation on every particle.

To conclude, the stability of the flat hexagonal configuration in
Fig.~\ref{fig1d} is confirmed by combination of analytical and numerical
approaches.

\section*{Appendix B: More information about the relaxation process and the
bond-angle distributions in fully relaxed lattices}

In this appendix, we present more information about the relaxation of truncated
hexagonal membranes and the variation of the lattice shape in the relaxation
process. We also present the distributions of the bond
angle along the layers of the lattice in the fully relaxed lattices at varying $k_e$.

In the main text, the stretched planar shape has been identified as the lowest-energy
state of the hexagonal membrane under the long-range electrostatic force. To check if the
flatness of the membrane is related to the hexagonal shape, we further discuss the
relaxation of truncated hexagonal membranes. Specifically, we remove $n_r$ rows of
particles from two opposite edges in the hexagon-shaped lattice, and obtain truncated
hexagonal lattices with broken $C_6$ symmetry as shown in Fig.~\ref{fig4.1}. These
elongated lattices are fully relaxed by the same procedure as in the main text.

We systematically check truncated hexagonal membranes of varying size in the
broad parameter space of $k_e$ and $k_b$: $k_e=\{0.001, 0.01, 0.1, 1, 10,
100\}$, and $k_b=\{0.01, 0.1, 1, 10, 100\}$. $n=\{2, 3, 4, 5, 6, 7, 8, 9\}$. For
each given value of $n$, $n_r=\{1,\ldots,n-1\}$. It turns out that the truncated
membranes uniformly converge to the flat shape in all of the cases. Some typical
cases are shown in Fig.~\ref{fig4.1}.  The value of the dimensionless quantity
$\tilde{h}$, which is introduced to characterize the degree of the out-of-plane
deformation in the main text, is within the order of $10^{-3}$ in all of the
cases in our simulations. To conclude, this observation clearly shows that the
flat lowest-energy shapes are uncorrelated to the hexagonal shape of the
membrane.

Furthermore, we track the variation of the hexagonal lattice shape in the relaxation process
from the perspective of the integral of the Gaussian curvature. For a smooth
surface $\Omega$ with piecewise smooth boundary consisting of $p$ regular curves
denoted as $C_1...C_p$, according to the generalized Gauss-Bonnet
theorem~\cite{do1992riemannian}, the integral of the Gaussian curvature is:
\begin{equation} \label{KG1}
	\int_{\Omega} K_G dA = 2\pi\chi(\Omega) - \sum_{i=1}^{p}\int_{C_i}k_g ds - \sum_{i=1}^{p}\gamma_i,
\end{equation}
where $\chi(\Omega)$ is the Euler characteristic of the surface $\Omega$.  $k_g$
is the geodesic curvature. $\gamma_i$ is the turning angle from curve $C_i$ to
$C_{i+1}$ at their meeting point; $C_{p+1} = C_1$. 

Note that within our numerical precision the following expression is verified to
be an invariant in the relaxation of the lattice:
\begin{equation} \label{KG_v}
	\int_{\Omega} K_G dA +  \sum_{i=1}^{p}\gamma_i.
\end{equation}
The temporally varying values of the first and the second terms for the case
of $n=6$ are presented in Table I.  It is found that the value of the
expression in Eq.~(\ref{KG_v}) is very close to $2\pi$; $2\pi \approx
6.2831853$. In comparison with Eq.~(\ref{KG1}), the integral of the geodesic
curvature is vanishingly small in the entire relaxation process of the lattice.
As such, Eq.~(\ref{KG1}) is verified within our numerical precision.

\begin{table*}[!ht] 
\centering
\renewcommand{\arraystretch}{1.2}
\label{tb1} 
\begin{tabular}{c|c|p{1.4cm}<{\centering}|p{1.4cm}<{\centering}|c|p{1.5cm}<{\centering}|p{1.5cm}<{\centering}|p{1.5cm}<{\centering}|c|c} 
	\hline 
	Number of steps ($\times10^{4}$)& 0 & 3 & 5 & 19 & 61 & 89 &	117 & 138 &	180 \\ 
	\hline
	$\int_{\Omega}K_G dA$& 0 & 4.589743 & 3.885128 & 1.352282 &  -3.802723 & -1.881432 & -0.658368 & -0.078556 & 0.000004 \\ 
	\hline 
	$\sum_{i}\gamma_i$& 6.283185 & 1.693442 & 2.398058 & 4.930903 &  10.085908 & 8.164617 & 6.941553 & 6.361742 & 6.283181 \\ 
	\hline 
	$\int_{\Omega}K_G dA+\sum_{i}\gamma_i$& 6.283185 & 6.283185 & 6.283186 &  6.283185 & 6.283185 & 6.283185 & 6.283185 & 6.283186 & 6.283185 \\ \hline
\end{tabular}
  \caption{Variations of the integral of the Gaussian curvature
  ($\int_{\Omega}K_G dA$) and the sum of the turning angles ($\sum_{i}\gamma_i$)
  in the relaxation of the lattice. The numbers in the bottom line are very close to
  $2\pi$; $2\pi \approx 6.2831853$. The generalized Gauss-Bonnet theorem is thus
  verified within our numerical precision. $n=6$. $k_b=0.01$. $k_e=100$. }
\end{table*}

Now we apply Eq.(\ref{KG1}) to our triangular lattice system, and obtain the
expression for the integral of the Gaussian curvature over the hexagonal cell
(denoted as $S_v$) surrounding the vertex $v$ as shown in Fig.~\ref{fig2a}:
\begin{equation} \label{KG_sv}
	\int_{S_v}K_G dA=2\pi - \sum_{i=0}^{k}\theta_i,
\end{equation}
where $\theta_i$ is the angle of the two adjacent bonds $vv_i$ and $vv_{i+1}$.
Note that $\chi(S_v)=1$, and $\sum_{i}\gamma_i=\sum_{i}\theta_i$. In the
derivation for Eq.(\ref{KG_sv}), it is assumed that the boundary curve is
geodesic. By the sum of the integrated Gaussian curvature in Eq.(\ref{KG_sv})
over the whole lattice, we obtain the total Gaussian curvature.

We systematically track the variation of the total Gaussian curvature in the relaxation
process for varying values of $n$, $k_e$ and $k_b$: $n=\{1, 2, 3, 4, 5, 6, 7, 8, 9\}$,
$k_b=\{0.01, 0.1, 1, 10, 100\}$ and $k_e=\{0.001, 0.01, 0.1, 1, 10, 100\}$. The main
results are summarized in Fig.~\ref{fig2b}. We see that the total Gaussian
curvature uniformly converges to zero in all of the cases. Oscillation of the
curve occurs in the relaxation process, indicating the exploration of the system
into both positively and negatively curved shapes (see the inset figure).
Closer examination of 54 oscillating curves shows the preference of negatively
curved shapes (46 out of 54) in the final stage of the relaxation process. This
observation implies that thermal fluctuation tends to deform the membrane to the
hyperbolic shape. The inset figures also show the smoothening of the shape with
the reduction of temperature.

\begin{figure*}[t]  
	\centering 
	\subfloat[$k_e=0.01$]{\includegraphics[width = .28\linewidth]{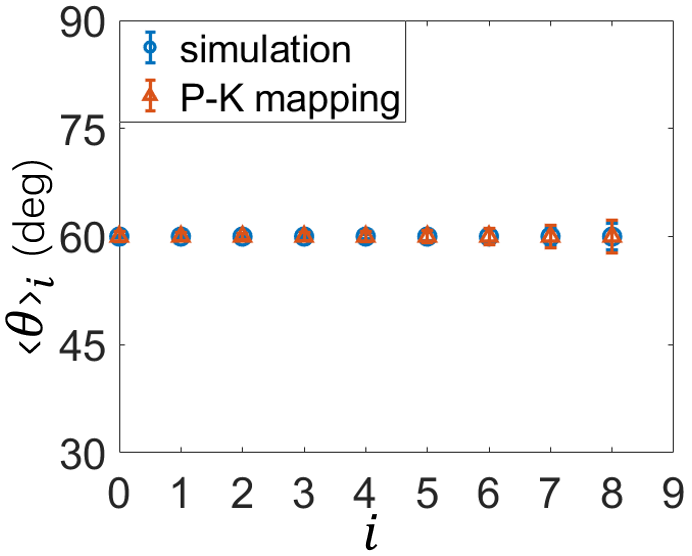}%
		\label{fig4.2a}}
	\subfloat[$k_e=1$]{\includegraphics[width = .28\linewidth]{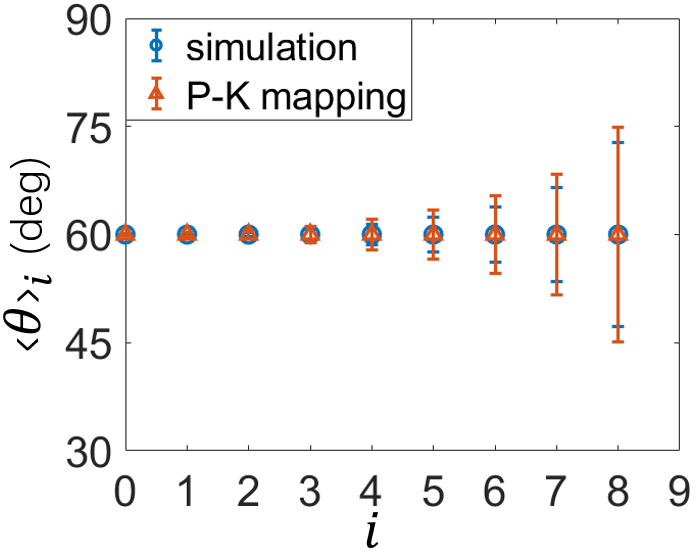}%
		\label{fig4.2b}}
	\subfloat[$k_e=100$]{\includegraphics[width = .28\linewidth]{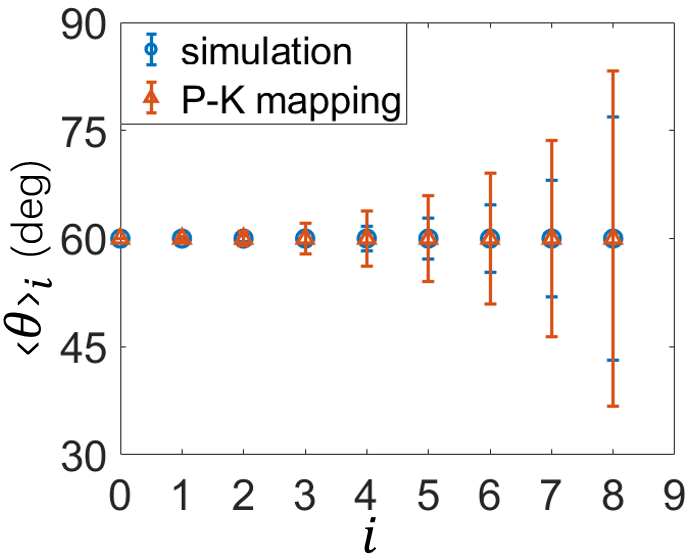}%
		\label{fig4.2c}}
  \caption{Distributions of the average bond angle $\langle\theta\rangle_i$
  in each layer $i$ for both the lowest-energy configuration obtained by
  simulations (blue circle) and the configuration generated by the
  Poincar$\rm{\acute{e}}$-Klein mapping (orange triangle) at varying $k_e$. The
  averaging procedure in each layer is over all of the bond angles associated
  with each vertex along the layer. The magnitude of the standard deviation is
  indicated by the length of the error bars. The error
  bars associated with simulations are slightly shorter than those associated
  with the Poincar$\rm{\acute{e}}$-Klein mapping for large values of $n$ in (b)
  and (c). $n=9$. $k_b=0.01$. The optimal
  values for the parameters $\lambda$ and $\Gamma$ in the Poincar$\rm{\acute{e}}$-Klein
  mapping, as well as the value for $L$, are listed here: (a) $\lambda = 0.010$,
  $\Gamma = 50.311$, $L = 0.002$. (b) $\lambda = 0.010$, $\Gamma = 51.541$, $L=
  0.004$. (c) $\lambda = 0.004$, $\Gamma = 129.374$, $L = 0.011$. }
	\label{fig4.2}
\end{figure*}

In Fig.~\ref{fig4.2}, the distributions of the bond angle along the layers of
the lattice in the lowest-energy states at varying $k_e$ are presented. We see
that the average bond angle is uniformly $60$ degrees as the value of $k_e$ is
varied from $k_e=0.01$ (a), $k_e=1$ (b), and $k_e=100$ (c). It indicates that
the preserved average bond angle is a common feature in the
electrostatics-driven deformation of the lattice. In the central region of the
lattice, the bond angle is well preserved in the deformation. We also notice
that the magnitude of the standard deviation in the bond-angle distribution
increases as the value of $k_e$ is increased.

\section*{Appendix C: Elasticity analysis of elastic disk upon stretching}	

In this appendix, we present analytical continuum elasticity analysis of an
isotropic elastic disk by applying a radial outward force on the boundary.

The distribution of stress over the disk in mechanical equilibrium is governed
by the following equation of equilibrium~\cite{Landau1986}:
\begin{equation}\label{eq4.3}
	\partial_{k}\sigma_{ik}=0, \qquad i,k = x,y.
\end{equation}
In the following, we rewrite Eq.(\ref{eq4.3}) in terms of the displacement
vector $\mathbf{u(\mathbf{x})}$.

The stress-strain relation is given by 
\begin{equation}\label{eq4.1}
	\left\{
	\begin{aligned}
		\sigma_{xx}&=\frac{E}{1-\sigma^2}(u_{xx}+\sigma u_{yy}),\\
		\sigma_{yy}&=\frac{E}{1-\sigma^2}(u_{yy}+\sigma u_{xx}),\\
		\sigma_{xy}&=\frac{E}{1+\sigma}u_{xy}.\\
	\end{aligned}
	\right.
\end{equation}
where $\sigma_{xx}$, $\sigma_{yy}$, $\sigma_{xy}$ are the components of the stress
tensor; $u_{xx}$, $u_{yy}$, $u_{xy}$ are the components of the strain tensor; $E$ is
the Young's modulus and $\sigma$ is the Poisson's ratio. 
For small deformations, the strain tensor is given by
\begin{equation}\label{eq4.2}
  u_{xx}=\frac{\partial u_x}{\partial x}, u_{xy}=\frac{1}{2}(\frac{\partial
  u_x}{\partial y}+\frac{\partial u_y}{\partial x}), u_{yy}=\frac{\partial
  u_y}{\partial y}.
\end{equation}
where $u_x, u_y$ are the components of the displacement vector. 
Substituting Eq.~(\ref{eq4.1}) and (\ref{eq4.2}) into Eq.~(\ref{eq4.3}), we obtain
the balance equations in terms of the displacement vector:
\begin{equation}\label{eq4.4}
	\left\{
	\begin{aligned}
		\frac{1}{1-\sigma^2}\frac{\partial^2 u_x}{\partial x^2}+\frac{1}{2(1+\sigma)}\frac{\partial^2 u_x}{\partial y^2}+\frac{1}{2(1-\sigma)}\frac{\partial^2 u_y}{\partial x\partial y}&=0,\\
		\frac{1}{1-\sigma^2}\frac{\partial^2 u_y}{\partial y^2}+\frac{1}{2(1+\sigma)}\frac{\partial^2 u_y}{\partial x^2}+\frac{1}{2(1-\sigma)}\frac{\partial^2 u_x}{\partial x\partial y}&=0.\\
	\end{aligned}
	\right.
\end{equation}
Equations~\ref{eq4.4} can be written in the vector form		
\begin{equation}\label{eq4.5}
	\nabla(\nabla\cdot\mathbf{u(\mathbf{x})}) - \frac{1}{2}(1-\sigma)\nabla\times(\nabla\times\mathbf{u(\mathbf{x})})=0,
\end{equation} 
where $\mathbf{u(\mathbf{x})}$ is the displacement vector at the point
$\mathbf{x}$ on the undeformed disk.

Upon the radial outward force $f$ on the boundary, the disk is subject to radial
stretching. We therefore search for the solution of rotational symmetry; i.e.,
the displacement vector $\mathbf{u(\mathbf{x})}$ is independent of the polar
angle $\phi$. Furthermore, the azimuthal component of the displacement vector is
zero. In other words, the solution is in the form of
$\mathbf{u(\mathbf{x})}=u_{r}(r)\hat{r}$. Here, we work in the polar coordinates
$(r,\phi)$. Equation~(\ref{eq4.5}) becomes 
\begin{equation}\label{eq4.6}
	\nabla\cdot\mathbf{u} =\frac{1}{r}\frac{\mathrm{d} (ru_r)}{\mathrm{d} r}=constant. 
\end{equation} 	
We therefore have 
\begin{equation}\label{eq4.7}
  u_r=C_1 r+\frac{C_2}{r},
\end{equation}
where the constants $C_1$ and $C_2$ are to be determined by the boundary
condition. By applying the boundary conditions that the displacement is finite
at $r=0$ and $\sigma_{rr}=f$ at $r=R$, we finally obtain the expressions for the strain field:
\begin{equation}\label{eq4.10}
  u_{rr}= u_{\phi\phi}=\frac{1-\sigma}{E}f,\ \  u_{r\phi}=0.
\end{equation}
Equation~(\ref{eq4.10}) shows that the strain field created by the boundary
radial force is homogeneous.

\section*{Appendix D: On the connection of the Poincar$\rm{\acute{e}}$ disk and the Klein disk}

In this appendix, we will show that under the Poincar$\rm{\acute{e}}$-Klein mapping the line element
$\mathrm{d}s_p$ in the Poincar$\rm{\acute{e}}$ disk is identical to the line element
$\mathrm{d}s_k$ in the Klein disk.

Consider an arbitrary point $(x_p, y_p)$ in the Poincar$\rm{\acute{e}}$ disk. Under the
Poincar$\rm{\acute{e}}$-Klein mapping~\cite{benedetti2012lectures}
\begin{equation}\label{eq3.7}
	f(z)=\frac{2z}{1+|z|^2},\quad |z|\leqslant 1,
\end{equation} 
it is mapped to $(x_k, y_k)$:
\begin{equation}\label{eq4.11}
	x_k=\frac{2x_p}{1+x_p^2+y_p^2},\qquad y_k=\frac{2y_p}{1+x_p^2+y_p^2}.
\end{equation}
From Eqs.(\ref{eq4.11}), we have
\begin{equation}\label{eq4.12}
\begin{aligned}
    &\mathrm{d}x_k=\frac{2(1-x_p^2+y_p^2)}{(1+x_p^2+y_p^2)^2}\mathrm{d}x_p-\frac{4x_p
    y_p}{(1+x_p^2+y_p^2)^2}\mathrm{d}y_p,\\ &\mathrm{d}y_k=-\frac{4x_p
    y_p}{(1+x_p^2+y_p^2)^2}\mathrm{d}x_p+\frac{2(1+x_p^2-y_p^2)}{(1+x_p^2+y_p^2)^2}\mathrm{d}y_p.\\
\end{aligned}
\end{equation}
Therefore, the line element $\mathrm{d}s_k$ in the Klein disk is 
\begin{equation}\label{eq4.13}
	\mathrm{d}s_k^2= \frac{\mathrm{d}x_k^2+\mathrm{d}y_k^2}{1-x_k^2-y_k^2}+\frac{(x_k\mathrm{d}x_k+y_k\mathrm{d}y_k)^2}{(1-x_k^2-y_k^2)^2}.
\end{equation}
By inserting Eqs.~(\ref{eq4.11}) and (\ref{eq4.12}) into Eq.~(\ref{eq4.13}), we have
\begin{equation}\label{eq4.14}
	\mathrm{d}s_k^2=\frac{4(\mathrm{d}x_p^2+\mathrm{d}y_p^2)}{(1-x_p^2-y_p^2)^2}.
\end{equation}
The term in right hand side of Eq.(\ref{eq4.14}) is recognized as
$\mathrm{d}s_p^2$. Therefore, we have shown that
$\mathrm{d}s_k^2=\mathrm{d}s_p^2$ under the Poincar$\rm{\acute{e}}$-Klein
mapping as defined in Eq.(\ref{eq3.7}). This result is as expected, because both
the Poincar$\rm{\acute{e}}$ disk and the Klein disk represent the common
hyperbolic plane.

\section{Acknowledgements}

This work was supported by the National Natural Science Foundation of China
(Grants No. BC4190050).


\end{document}